\documentclass[pmlr]{jmlr} % This is the PMLR single-column format

\usepackage{booktabs} % For better tables

\jmlrvolume{312}
\jmlryear{2026}
\jmlrworkshop{AAAI 2026 Workshop on Audio-AAAI}
\editors{
Tatsuya Komatsu,
Keisuke Imoto, 
Xiaoxue Gao,
Nobutaka Ono,
Nancy F. Chen
}

\title{AudioRAG: A Challenging Benchmark for Audio Reasoning and Information Retrieval}

\author{%
  \Name{Jingru Lin} \Email{jingrulin@u.nus.edu}\\
  \addr Department of Electrical and Computer Engineering, \\
  National University of Singapore, Singapore\\
  \AND
  \Name{Chen Zhang} \Email{e0397123@u.nus.edu}\\
  \addr Department of Electrical and Computer Engineering, \\
  National University of Singapore, Singapore\\
  \AND
  \Name{Tianrui Wang} \Email{wangtianrui@tju.edu.cn}\\
  \addr College of Intelligence and Computing, \\
  Tianjin University, China\\
  \AND
  \Name{Haizhou Li} \Email{haizhou.li@nus.edu.sg}\\
  \addr Department of Electrical and Computer Engineering, \\
  National University of Singapore, Singapore\\
  \addr SRIBD, School of Data Science, \\
  The Chinese University of Hong Kong, Shenzhen, Guangdong\\
}

\begin{document}

\maketitle

\begin{abstract}
Due to recent advancements in Large Audio-Language Models (LALMs) that demonstrate remarkable performance across a range of sound-, speech- and music-related tasks, there is a growing interest in proposing benchmarks to assess these models. Existing benchmarks generally focus only on reasoning with internal knowledge, neglecting real-world scenarios that require external information grounding. To bridge this gap, we introduce AudioRAG, a novel benchmark designed to evaluate audio-based reasoning augmented by information retrieval in realistic web environments. This benchmark comprises both LLM-generated and manually curated question-answer pairs. Our evaluations reveal that even the state-of-the-art LALMs struggle to answer these questions. We therefore propose an agentic pipeline that integrates audio reasoning with retrieval-augmented generation, providing a stronger baseline for future research. 
\end{abstract}


\section{Introduction}

\label{sec:intro}
Advancements in Large Audio-Language Models (LALMs) highlight their potential to unify speech, sound and music understanding within a single multimodal framework~\cite{chu2024qwen2,kong2024audio,ghosh2025audio}. These advanced LALMs exhibit impressive performance across a wide range of understanding and generation tasks involving speech, sound, and music, such as speech recognition~\cite{ma2025speech}, music genre classification~\cite{meguenani2025music}, audio captioning~\cite{chen2025slam}, etc. While these tasks assess foundational audio understanding, they largely emphasize perceptual recognition rather than complex reasoning, which characterizes more sophisticated forms of intelligence.

Recently, there has been a growing research interest in exploring the capacity of LALMs to perform multi-hop or deep reasoning, and several benchmarks have been proposed for this purpose~\cite{yang25g_interspeech,ma2025mmar,sakshimmau}. 
For example, SAKURA~\cite{yang25g_interspeech} evaluates multi-hop reasoning capability. MMAR~\cite{ma2025mmar} evaluates the LALMs' deep reasoning capabilities, including expert-level perceptual understanding and multi-step inference. However, these benchmarks primarily assess reasoning over internal, parameterized knowledge only. 

In real-world scenarios, users often ask questions that extend beyond a model's internal knowledge. For example, a user might inquire about the final score of a soccer match mentioned in a recent news broadcast. If such information is absent from the model's training data, the model may generate an unfaithful response. This problem is often recognized as \textit{hallucination} in large language model (LLM) research~\cite{huang2025survey}. To alleviate \textit{hallucination}, retrieval-augmented generation (RAG) has been widely adopted, as it grounds the model outputs in retrieved and verifiable knowledge~\cite{izacard2023atlas,gao2023retrieval}. By coupling generation with retrieval, RAG effectively enhances factual accuracy and reduces reliance on internal memorization. Nevertheless, existing audio reasoning benchmarks have not yet accounted for scenarios where audio-based reasoning is combined with retrieval. 

To address this gap, we introduce AudioRAG, a new benchmark featuring challenging questions tailored for audio reasoning with information seeking under real-world web environments. AudioRAG is constructed based on both open-sourced datasets and human-collected data. For open-sourced datasets, we take one attribute of the audio, e.g., genre of the music, species of the animal, from the metadata and prompt LLMs to generate questions based on that attribute. The generated questions are designed to require both audio processing and information retrieval steps. To further diversify the dataset, we manually collect videos from online sources and extract their audio tracks. Based on the extracted audios, we curate multi-hop questions similar to the ones generated by LLMs from the open-sourced datasets. 

Based on the benchmark, we evaluate several state-of-the-art LALMs. Our results show that existing LALMs struggle to answer these audio-based multi-hop reasoning questions. This observation is consistent with findings from previous studies~\cite{ma2025mmar,yang25g_interspeech}. Moreover, when lacking the necessary knowledge to answer the question, the LALMs provide hallucinated responses. To better address such challenges, we propose an agentic pipeline capable of performing both audio reasoning and information retrieval, serving as a strong baseline for future research.

Overall, our contributions are: (1) Introducing AudioRAG, the first benchmark for systematically evaluating models' multi-hop audio reasoning and information retrieval capabilities; (2) We evaluate LALMs on the benchmark, revealing that current LALMs struggle to answer the questions; (3) We develop an agentic pipeline that shows a relative improvement of up to 24.9\%. 

\begin{figure*}[t]
    \caption{Data construction process (left) and the agentic pipeline (right).}
    \includegraphics[width=\linewidth]{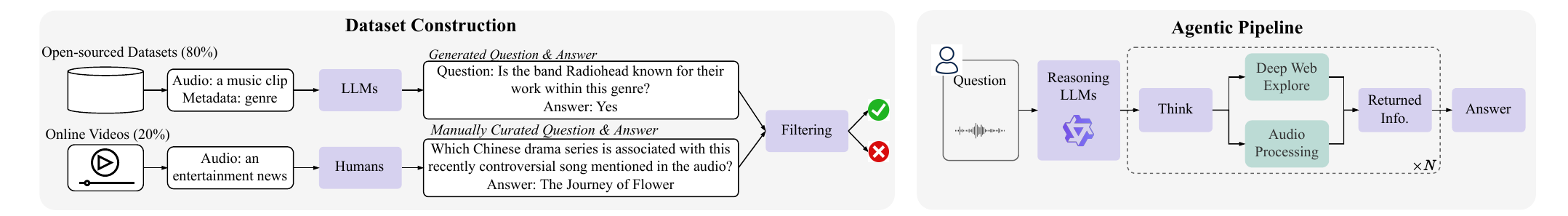}
\label{fig:pipeline}
\end{figure*}

\section{Methods}

Each question in our dataset consists of a multi-hop question based on an audio context and the corresponding answer. The question is either in text or audio format, while the context is always in audio format. The multi-hop question is either generated based on open-source datasets or manually curated by humans. The following sections give details about the constructed dataset.

\subsection{Dataset Statistics}
% data sources
We first examine many open-source datasets and select the following for our dataset construction: 
\begin{itemize}
    \item \textbf{Speech/Sound/Music}: MMAU~\cite{sakshimmau}, CinePile~\cite{rawal2024cinepile}, \\ Multitask-National-Speech-Corpus~\cite{wang2025advancing}, FMA~\cite{fma_dataset}, Jazznet~\cite{adegbija2023jazznet}, MusicNet~\cite{thickstun2017learning}, iNaturalist~\cite{chasmai2025inaturalistsoundsdataset}
    \item \textbf{Text}: CHEER~\cite{buzz2025huang}
\end{itemize}
% We select these datasets based on the possibility of forming multi-hop questions that involve at least one audio processing step and one information retrieval step. For text datasets, we use index-tts to synthesise speech context.
% b) if the questions are reflective of true user questions.

On top of these open-source datasets, we also collect additional audio from online sources, then manually curate multi-hop questions based on the collected audio. 
% In addition, we believe that audio is an important source of information for humans, and yet, with the rise of media and generative AI, it is important to verify the authenticity of the information. For this purpose, we also create some audio that might contain fake information. 

Overall, we collect and generate 500 samples and release them at \url{https://github.com/jingru-lin/AudioRAG}\footnote{Details about the original sample sources are also provided.}.

% average audio lengths

\begin{table}[t]
\centering
\caption{Prompts for the LLM to generate the multi-hop questions. Here, \texttt{[ATTRIBUTE]} refers to an attribute of the audio, such as the genre of music, the source of a sound etc.}
\begin{tabular}{p{0.95\linewidth}}
\hline
A prompt example to generate the multi-hop questions \\
\hline
You are good at creating multi-hop questions. Your task is to generate multi-hop questions based on the given information about [ATTRIBUTE].

**Requirements**: \\
1. Do not mention the given \texttt{[ATTRIBUTE]} directly. You should assume an audio is given and the \texttt{[ATTRIBUTE]} must be derived from the audio. You can refer to it indirectly (e.g., using pronouns). \\
2. If possible, create a question that might involve an information-retrieval step to answer the question. \\
3. The final question must arrive at a clear correct answer. \\
4. Phrase the question naturally so it reads like a real user query. \\
5. Provide an answer to the questions. \\
6. No questions about the recognition of \texttt{[ATTRIBUTE]} should be generated. \\
Examples: \\
\texttt{[FEW-SHOT\_EXAMPLES]} \\
\hline
\end{tabular}
\label{tab:llm_prompts}
\end{table}

\subsection{Question Generation}
We utilize GPT-4o to generate the multi-hop question-answer pairs. Specifically, we take one attribute of the audio from metadata of the open-source datasets, for example, genre of music, species of animals heard in the audio, transcriptions of speech, and prompt GPT-4o to generate multi-hop questions that might require external knowledge to answer. Few-shot examples are provided in the prompts. Different templates and few-shot examples are needed for the audio samples from different datasets. Here, we only give an example of the prompt to illustrate the idea, as shown in Table~\ref{tab:llm_prompts}. The generated question should resemble a true user question after hearing the audio. Figure~\ref{fig:pipeline} shows an example.
% Some examples are shown in Table~\ref{tab:question_examples}.

In addition, we manually collect some data from online sources. These questions contain timely information and are less likely to be exposed to data contamination. Therefore, current models lack sufficient parametric knowledge about the information required to answer the question, which makes retrieving external information highly important. An example question is shown in Figure~\ref{fig:pipeline}.

% , where the content may already be included in the model pretraining

\subsection{Data Filtering}
We employ two filters: a) Question Validity Filter; b) Answer Correctness Filter. The first ensures the quality of the generated questions. Specifically, we employ both LLM and annotators to verify whether the question has one unique answer. Questions lacking a unique answer are removed. 
The second filter focuses on verifying the answer correctness. To ensure the correctness of answers generated, we employ LLM-based agents with a search tool that allows for up-to-date information retrieval when answering the questions. 
We provide the ground-truth audio attributes (text) that were used to generate the questions for agents to answer the question. In such a way, it reduces the noise introduced when processing the audio, and the agents can focus exclusively on answer correctness. 
% Since we know the attribute of the audio that is used to generate the questions, we can turn the questions into pure multi-hop text-based questions. 
One example is shown in Appendix~\ref{appendix:data_filter}. 
% We then employ the advanced LLM-based agents to verify the answers.
When the agents' answers differ from the originally generated ones, human annotators review those cases and the corresponding questions are either revised or discarded.

\section{Agentic Pipeline}

Our initial results show that the most open-source Large Audio Language Models (LALMs) struggle to answer these multi-hop questions. There are two main reasons. Firstly, the LALMs, while capable of handling audio data, show weak text capabilities, including instruction following, question interpretation, etc. Therefore, most LALMs struggle to break down the non-straightforward multi-hop questions into solvable subtasks. Secondly, the LALMs lack the knowledge, especially timely information, to answer the question. To tackle these problems, we propose an LLM-based agentic pipeline. Our pipeline makes use of a text-based LLM to handle user queries and drive tool usage. This pipeline is integrated with two tools: an audio processing tool and a search tool. While the former compensate for the text capabilities that the large language models lack, the latter complement the model with external knowledge. We will introduce the details of the pipeline in the following sections.

\subsection{Pipeline Details}
Our pipeline is based on WebThinker~\cite{Li2025WebThinker}, but we enhance it with audio-processing capabilities. 
Given a user query, consisting of a question $q$ that is either in text or audio form and an audio context $c$, the pipeline generates a solution for answering the user's query regarding the audio context, guided by an instruction $I$. To answer the complex, multi-hop queries, the pipeline implements an autonomous \textbf{Think-Call-Answer} strategy, which iteratively thinks upon the current status, calls relevant tools, and generates intermediate/final responses. 
The final solution comprises a reasoning chain $\mathcal{R}$ and a final output $y$. The reasoning LLM in the pipeline orchestrates the whole process and autonomously invokes tools from an available set $\mathcal{T}$ during its reasoning process. The whole process can be formalized as:

\begin{align}
P(R, y \mid I, q, \mathcal{T}) 
&= \underbrace{\prod_{t=1}^{T_r} 
   P(R_t \mid R_{<t}, I, q, \{O_\tau\}_{\tau < t})}_{\text{Reasoning with Tools}} \notag \\
&\quad \cdot 
   \underbrace{\prod_{t=1}^{T_y} 
   P(y_t \mid y_{<t}, R, I, q)}_{\text{Final Output Generation}},
\end{align}
where $\mathcal{T}_r$ is the number of tokens in the reasoning sequence $\mathcal{R}$. The token at position $t$ is $\mathcal{R}_t$, and $\mathcal{R}_{<t}$ represents all tokens generated before position $t$.  
Similarly, $T_y$ is the length of the output sequence $y$, with $y_t$ being the token at position $t$ and $y_{<t}$ indicating all generated output tokens before position $t$.
$\{ O_\tau \}_{\tau < t}$ denotes the outputs of all tool calls
made before position t. The tool set consists of two tools $\mathcal{T} = \{\mathcal{T}_{\text{exp}}, \mathcal{T}_a\}$. $\mathcal{T}_{\text{exp}}$ is a deep web explorer tool that can retrieve information online and $\mathcal{T}_a$ is an audio processing tool that can extract necessary information from the $c$. During the reasoning process, the reasoning LLM needs to iteratively call one of the tools and generate intermediate outputs. 

\subsection{Tools}
When the LLM's internal knowledge is not enough to answer the user query, it can invoke the search tool $\mathcal{T}_{\text{exp}} \in \mathcal{T}$ to extract external knowledge. The deep web explorer tool is directly adopted from the WebThinker~\cite{Li2025WebThinker}, which conducts search actions and web browsing actions\footnote{Interested audiences may refer to the original paper for details.}. To extract relevant information from the audio, an audio processing tool is needed. 
% The reasoning LLM is prompted to generate instruction in the format $<$begin\_audio\_analysis$>$..$<$end\_audio\_analysis$>$. 
When an audio processing step is needed, the reasoning LLM will generate an audio-related query that extracts information from the audio. The query is wrapped in $<$begin\_audio\_analysis$>$[AUDIO-RELATED QUERY]$<$end\_audio\_analysis$>$. Upon receiving the query, the pipeline will initiate the audio processing tool, which will then output a response $O_a$ based on the instruction.

\section{Experiments}
\subsection{Raw Models}
We evaluate several open- and closed-sourced models, including Qwen2.5-Omni~\cite{xu2025qwen2}, Audio Flamingo 3~\cite{goel2025audio}, Audio-Reasoner~\cite{xie2025audio}, Baichuan-Omni~\cite{li2025baichuan}, Qwen3-Omni~\cite{xu2025qwen3} and Gemini-2.5-Flash~\cite{comanici2025gemini}. 

\subsection{Agentic Pipeline}
We use Qwen3-8B~\cite{yang2025qwen3} as the reasoning LLM to drive the whole pipeline. The search tool $\mathcal{T}_{\text{exp}}$ used is the Google search engine API. The audio processing tool is Qwen2.5-Omni~\cite{xu2025qwen2} / Qwen3-Omni~\cite{xu2025qwen3}. Both reasoning LLM and audio processing tool are served with vLLM~\cite{kwon2023efficient}, with four A100 (40GB) GPUs each. 

\subsection{Main Results}

\begin{table}[t]
\centering
\caption{Main results. Since the agentic pipeline consists of audio and text LLMs, the size is broken down into size of (audio) + (text). The size of Gemini-2.5-Flash is unknown as it is a closed-source model. The rest are open-source models.}
\begin{tabular}{ccc}
\hline
Model & Size & Accuracy (\%) \\
\hline
\hline 
\multicolumn{3}{c}{\textbf{Raw Model}} \\
\hline
Qwen2.5-Omni~\cite{xu2025qwen2} & 7B & 32.2 \\
Audio Flamingo 3~\cite{goel2025audio} & 8.3B & 28.8  \\
Audio-Reasoner~\cite{xie2025audio} & 8.4B & 20.2 \\
Baichuan-Omni~\cite{li2025baichuan} & 11B & 24.4 \\
Qwen3-Omni~\cite{xu2025qwen3} & 30B & 37.0 \\
Gemini-2.5-Flash~\cite{comanici2025gemini} & - & \textbf{45.0} \\
\hline
\hline
\multicolumn{3}{c}{\textbf{Agentic Pipeline}} \\
\hline
Qwen2.5-Omni + Qwen3-8B~\cite{yang2025qwen3} & 7B + 8B & 39.5 \\
Qwen3-Omni + Qwen3-8B~\cite{yang2025qwen3} & 30B + 8B & \textbf{46.2} \\
\hline
\end{tabular}
\label{tab:main_results}
\end{table}

Table~\ref{tab:main_results} shows the performance of both raw models and agentic pipelines on the benchmark. The reported metric is accuracy (\%), which records to number of correct questions answered. Each question is provided with an answer and GPT-4o is used to evaluate the generated answer. The evaluation prompt is shown in Appendix~\ref{appendix:eval_prompt}. The results in Table~\ref{tab:main_results} are averaged over three runs. 

From Table~\ref{tab:main_results}, the strongest raw model is the closed-source Gemini-2.5-Flash, achieving 45\% accuracy, followed by open-sourced Qwen3-Omni with 37.0\%. All other open-sourced models perform notably worse, highlighting a clear performance gap between open-source and closed-source models. This is likely due to the latter's stronger internal knowledge or agentic abilities. Among all open-sourced models, the Qwen family consistently outperforms others, demonstrating stronger audio reasoning capabilities.

Compared to the raw models, the agentic pipeline consistently delivers superior performance. Specifically, Qwen2.5-Omni and Qwen3-Omni achieve accuracies of 32.2\% and 37.0\%, respectively. When integrated into the agentic pipeline with Qwen3-8B as the reasoning model, their accuracies rise to 39.5\% and 46.2\%, corresponding to relative improvements of 22.7\% and 24.9\%. 

\begin{table}[t]
\centering
\caption{Case study comparing outputs from a raw model and agentic pipeline. Reasoning error is highlighted in red.}
\begin{tabular}{p{0.95\linewidth}}
\hline
Audio: an audio clip about ``The Wizard of Oz'' \\
Question: How is the movie mentioned in this audio related to Wicked?\\
\hline
\underline{Answer by Qwen3-Omni}: \\
The movie mentioned is **The Wizard of Oz**. \textcolor{red}{The audio does not mention the movie **Wicked** at all.}

\underline{Answer by Agentic Pipeline with Qwen3-Omni}: \\
Wicked is a prequel to The Wizard of Oz, exploring... \\

\underline{Ground-truth Answer}: \\
Wicked is a prequel to The Wizard of Oz \\
\hline
\end{tabular}
\label{tab:case_study}
\end{table}

Table~\ref{tab:case_study} shows an example of actual outputs from a raw model and an agentic pipeline. The question in this example requires reasoning beyond the information directly available in the audio content. The raw model focuses purely on the audio input, correctly identifying the mentioned movie but failing to infer the link between the two movies. In contrast, the agentic pipeline demonstrates a clearer understanding of the question and effectively retrieves external knowledge to identify the correct relationship between the two movies.

Table~\ref{tab:main_results} and~\ref{tab:case_study} both show that the agentic pipeline bridges the gap in reasoning and knowledge integration between raw models and human-like multimodal understanding. While the raw models demonstrate perceptual capabilities, their performance is limited by a lack of contextual reasoning and external knowledge grounding. The agentic pipeline addresses these limitations.

\begin{figure}[t]
    \caption{Error types breakdown for incorrect answers. The x-axis is the category of errors. A refers to Reasoning Error, B refers to Audio Processing Error, C refers to Knowledge Error and D refers to Invalid Answer.}
  \includegraphics[width=\linewidth]{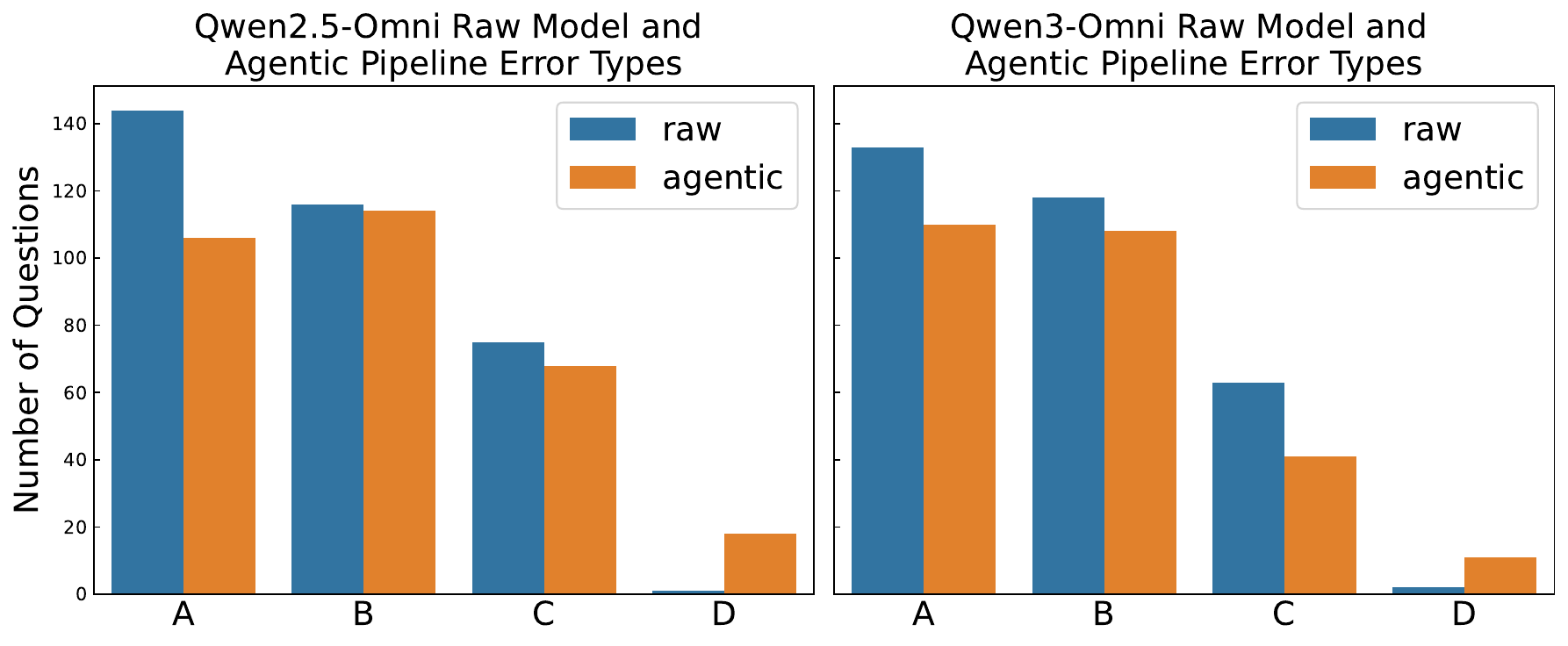}
\label{fig:error}
\end{figure}

\subsection{Analysis on Error Types}
In this section, we analyze the different error types made by the raw models and the agentic pipelines. We summarise four error types: a) Reasoning Error, where the model fails to understand the question or apply correct reasoning to arrive at the answer; b) Audio Processing Error, where the question requires to recognise a specific attribute of the audio but the model wrongly recognise it; c) Knowledge Error, where the question requires external knowledge beyond what is in the audio but the model gives the wrong answer due to lack of or incorrect knowledge; d) Invalid Answer: the model gives no response or incomplete output. GPT-4o is used to analyze the answers and categorize the incorrect answers into one of these categories. Figure~\ref{fig:error} shows the number of questions belonging to each error category. The evaluation prompt is shown in Appendix~\ref{appendix:eval_prompt}. This plot is based on outputs from raw models Qwen2.5-7B and Qwen3-30B and their corresponding agentic pipeline coupled with Qwen3-8B. 

From Figure~\ref{fig:error}, we observe that the agentic pipeline reduces Reasoning, Audio Processing and Knowledge Error, but tends to produce more Invalid Answer. The largest reduction is in Reasoning Error. This indicates that raw models often struggle to interpret complex multi-hop questions but the structured multi-step reasoning of the agentic pipeline significantly improves comprehension. The second major improvement lies in Knowledge Error. This is due to the knowledge retrieval component in the agentic pipeline which enhances the model with up-to-date and factually accurate information. Improvements in Audio Processing Error are smaller, as the pipeline still relies on the raw model to process audio. Invalid Answer increases, likely because the more complex multi-step reasoning can lead the pipeline into infinite logical loops. 

\section{Conclusion}
In this paper, we introduce AudioRAG, the first benchmark designed to evaluate models' multi-hop audio reasoning and information retrieval capabilities. Our experiments show that current LALMs perform poorly on this challenging benchmark, while an agentic pipeline that integrates audio reasoning with information retrieval establishes a stronger baseline for future research.

\section{Acknowledgements}
This work is supported by Shenzhen Stability Science Program 2023, Shenzhen Key Lab of Multi-Modal Cognitive Computing, Shenzhen Science and Technology Program \\
ZDSYS20230626091302006 and the Program for Guangdong Introducing Innovative and Entrepreneurial Teams, Grant No. 2023ZT10X044.

\bibliography{aaai2026}

@inproceedings{sakshimmau,
  title={MMAU: A Massive Multi-Task Audio Understanding and Reasoning Benchmark},
  author={Sakshi, S and Tyagi, Utkarsh and Kumar, Sonal and Seth, Ashish and Selvakumar, Ramaneswaran and Nieto, Oriol and Duraiswami, Ramani and Ghosh, Sreyan and Manocha, Dinesh},
  year={2025},
  booktitle={The Thirteenth International Conference on Learning Representations},
}

@article{rawal2024cinepile,
  title={CinePile: A Long Video Question Answering Dataset and Benchmark},
  author={Rawal, Ruchit and Saifullah, Khalid and Basri, Ronen and Jacobs, David and Somepalli, Gowthami and Goldstein, Tom},
  journal={arXiv preprint arXiv:2405.08813},
  year={2024}
}

@article{wang2025advancing,
  title={Advancing Singlish Understanding: Bridging the Gap with Datasets and Multimodal Models},
  author={Wang, Bin and Zou, Xunlong and Sun, Shuo and Zhang, Wenyu and He, Yingxu and Liu, Zhuohan and Wei, Chengwei and Chen, Nancy F and Aw, AiTi},
  journal={arXiv preprint arXiv:2501.01034},
  year={2025}
}

@inproceedings{fma_dataset,
  title = {{FMA}: A Dataset for Music Analysis},
  author = {Defferrard, Micha\"el and Benzi, Kirell and Vandergheynst, Pierre and Bresson, Xavier},
  booktitle = {18th International Society for Music Information Retrieval Conference (ISMIR)},
  year = {2017},
  archiveprefix = {arXiv},
  eprint = {1612.01840},
  url = {https://arxiv.org/abs/1612.01840},
}

@inproceedings{adegbija2023jazznet,
  title={jazznet: A dataset of fundamental piano patterns for music audio machine learning research},
  author={Adegbija, Tosiron},
  booktitle={ICASSP 2023-2023 IEEE International Conference on Acoustics, Speech and Signal Processing (ICASSP)},
  pages={1--5},
  year={2023},
  organization={IEEE}
}

@inproceedings{thickstun2017learning,
    title={Learning Features of Music from Scratch},
    author = {John Thickstun and Zaid Harchaoui and Sham M. Kakade},
    year={2017},
    booktitle = {International Conference on Learning Representations (ICLR)}
}

@misc{chasmai2025inaturalistsoundsdataset,
      title={The iNaturalist Sounds Dataset}, 
      author={Mustafa Chasmai and Alexander Shepard and Subhransu Maji and Grant Van Horn},
      year={2025},
      eprint={2506.00343},
      archivePrefix={arXiv},
      primaryClass={cs.SD},
      url={https://arxiv.org/abs/2506.00343}, 
}

@inproceedings{buzz2025huang,
  title={Can Large Language Models Understand Internet Buzzwords Through User-Generated Content},
  author={Chen Huang and 
          Junkai Luo and 
          Xinzuo Wang and 
          Wenqiang Lei and 
          Jiancheng Lv},
  booktitle = {Annual Meeting of the Association for Computational Linguistics},
  year      = {2025}
}

@article{Li2025WebThinker,
  author       = {Xiaoxi Li and
                  Jiajie Jin and
                  Guanting Dong and
                  Hongjin Qian and
                  Yutao Zhu and
                  Yongkang Wu and
                  Ji{-}Rong Wen and
                  Zhicheng Dou},
  title        = {WebThinker: Empowering Large Reasoning Models with Deep Research Capability},
  journal      = {CoRR},
  volume       = {abs/2504.21776},
  year         = {2025},
  url          = {https://doi.org/10.48550/arXiv.2504.21776},
  doi          = {10.48550/ARXIV.2504.21776},
  eprinttype    = {arXiv},
  eprint       = {2504.21776},
  bibsource    = {dblp computer science bibliography, https://dblp.org}
}

@inproceedings{kwon2023efficient,
  title={Efficient Memory Management for Large Language Model Serving with PagedAttention},
  author={Woosuk Kwon and Zhuohan Li and Siyuan Zhuang and Ying Sheng and Lianmin Zheng and Cody Hao Yu and Joseph E. Gonzalez and Hao Zhang and Ion Stoica},
  booktitle={Proceedings of the ACM SIGOPS 29th Symposium on Operating Systems Principles},
  year={2023}
}

@inproceedings{yang25g_interspeech,
  title     = {{SAKURA: On the Multi-hop Reasoning of Large Audio-Language Models Based on Speech and Audio Information}},
  author    = {Chih-Kai Yang and Neo Ho and Yen-Ting Piao and Hung-yi Lee},
  year      = {2025},
  booktitle = {{Interspeech 2025}},
  pages     = {1788--1792},
  doi       = {10.21437/Interspeech.2025-839},
  issn      = {2958-1796},
}

@article{ma2025mmar,
  title={MMAR: A Challenging Benchmark for Deep Reasoning in Speech, Audio, Music, and Their Mix},
  author={Ma, Ziyang and Ma, Yinghao and Zhu, Yanqiao and Yang, Chen and Chao, Yi-Wen and Xu, Ruiyang and others},
  journal={arXiv preprint arXiv:2505.13032},
  year={2025}
}

@article{xu2025qwen2,
  title={Qwen2. 5-omni technical report},
  author={Xu, Jin and Guo, Zhifang and He, Jinzheng and Hu, Hangrui and He, Ting and Bai, Shuai and Chen, Keqin and Wang, Jialin and Fan, Yang and Dang, Kai and others},
  journal={arXiv preprint arXiv:2503.20215},
  year={2025}
}

@article{goel2025audio,
  title={Audio flamingo 3: Advancing audio intelligence with fully open large audio language models},
  author={Goel, Arushi and Ghosh, Sreyan and Kim, Jaehyeon and Kumar, Sonal and Kong, Zhifeng and Lee, Sang-gil and Yang, Chao-Han Huck and Duraiswami, Ramani and Manocha, Dinesh and Valle, Rafael and others},
  journal={arXiv preprint arXiv:2507.08128},
  year={2025}
}

@article{xie2025audio,
  title={Audio-reasoner: Improving reasoning capability in large audio language models},
  author={Xie, Zhifei and Lin, Mingbao and Liu, Zihang and Wu, Pengcheng and Yan, Shuicheng and Miao, Chunyan},
  journal={arXiv preprint arXiv:2503.02318},
  year={2025}
}

@article{li2025baichuan,
  title={Baichuan-Omni-1.5 Technical Report},
  author={Li, Yadong and Liu, Jun and Zhang, Tao and Chen, Song and Li, Tianpeng and Li, Zehuan and Liu, Lijun and Ming, Lingfeng and Dong, Guosheng and Pan, Da and others},
  journal={arXiv preprint arXiv:2501.15368},
  year={2025}
}

@article{xu2025qwen3,
  title={Qwen3-omni technical report},
  author={Xu, Jin and Guo, Zhifang and Hu, Hangrui and Chu, Yunfei and Wang, Xiong and He, Jinzheng and Wang, Yuxuan and Shi, Xian and He, Ting and Zhu, Xinfa and others},
  journal={arXiv preprint arXiv:2509.17765},
  year={2025}
}

@article{comanici2025gemini,
  title={Gemini 2.5: Pushing the frontier with advanced reasoning, multimodality, long context, and next generation agentic capabilities},
  author={Comanici, Gheorghe and Bieber, Eric and Schaekermann, Mike and Pasupat, Ice and Sachdeva, Noveen and Dhillon, Inderjit and Blistein, Marcel and Ram, Ori and Zhang, Dan and Rosen, Evan and others},
  journal={arXiv preprint arXiv:2507.06261},
  year={2025}
}

@article{yang2025qwen3,
  title={Qwen3 technical report},
  author={Yang, An and Li, Anfeng and Yang, Baosong and Zhang, Beichen and Hui, Binyuan and Zheng, Bo and Yu, Bowen and Gao, Chang and Huang, Chengen and Lv, Chenxu and others},
  journal={arXiv preprint arXiv:2505.09388},
  year={2025}
}

@article{chu2024qwen2,
  title={Qwen2-audio technical report},
  author={Chu, Yunfei and Xu, Jin and Yang, Qian and Wei, Haojie and Wei, Xipin and Guo, Zhifang and Leng, Yichong and Lv, Yuanjun and He, Jinzheng and Lin, Junyang and others},
  journal={arXiv preprint arXiv:2407.10759},
  year={2024}
}

@inproceedings{
  ghosh2025audio,
  title={Audio Flamingo 2: An Audio-Language Model with Long-Audio Understanding and Expert Reasoning Abilities},
  author={Ghosh, Sreyan and Kong, Zhifeng and Kumar, Sonal and Sakshi, S and Kim, Jaehyeon and Ping, Wei and Valle, Rafael and Manocha, Dinesh and Catanzaro, Bryan},
  booktitle={Forty-second International Conference on Machine Learning},
  year={2025},
  url={https://openreview.net/forum?id=xWu5qpDK6U}
}

@inproceedings{kong2024audio,
  title={Audio Flamingo: A Novel Audio Language Model with Few-Shot Learning and Dialogue Abilities},
  author={Kong, Zhifeng and Goel, Arushi and Badlani, Rohan and Ping, Wei and Valle, Rafael and Catanzaro, Bryan},
  booktitle={International Conference on Machine Learning},
  pages={25125--25148},
  year={2024},
  organization={PMLR}
}

@inproceedings{ma2025speech,
  title={Speech Recognition Meets Large Language Model: Benchmarking, Models, and Exploration},
  author={Ma, Ziyang and Yang, Guanrou and Yang, Yifan and Gao, Zhifu and Wang, Jiaming and Du, Zhihao and Yu, Fan and Chen, Qian and Zheng, Siqi and Zhang, Shiliang and others},
  booktitle={Proceedings of the AAAI Conference on Artificial Intelligence},
  volume={39},
  number={23},
  pages={24840--24848},
  year={2025}
}

@inproceedings{chen2025slam,
  title={Slam-aac: Enhancing audio captioning with paraphrasing augmentation and clap-refine through llms},
  author={Chen, Wenxi and Ma, Ziyang and Li, Xiquan and Xu, Xuenan and Liang, Yuzhe and Zheng, Zhisheng and Yu, Kai and Chen, Xie},
  booktitle={ICASSP 2025-2025 IEEE International Conference on Acoustics, Speech and Signal Processing (ICASSP)},
  pages={1--5},
  year={2025},
  organization={IEEE}
}

@inproceedings{meguenani2025music,
  title={Music genre classification using large language models},
  author={Meguenani, Mohamed El Amine and de Souza Britto, Alceu and Koerich, Alessandro Lameiras},
  booktitle={2025 IEEE Symposium on Computational Intelligence in Image, Signal Processing and Synthetic Media (CISM)},
  pages={1--7},
  year={2025},
  organization={IEEE}
}

@article{huang2025survey,
  title={A survey on hallucination in large language models: Principles, taxonomy, challenges, and open questions},
  author={Huang, Lei and Yu, Weijiang and Ma, Weitao and Zhong, Weihong and Feng, Zhangyin and Wang, Haotian and Chen, Qianglong and Peng, Weihua and Feng, Xiaocheng and Qin, Bing and others},
  journal={ACM Transactions on Information Systems},
  volume={43},
  number={2},
  pages={1--55},
  year={2025},
  publisher={ACM New York, NY}
}

@article{izacard2023atlas,
  title={Atlas: Few-shot learning with retrieval augmented language models},
  author={Izacard, Gautier and Lewis, Patrick and Lomeli, Maria and Hosseini, Lucas and Petroni, Fabio and Schick, Timo and Dwivedi-Yu, Jane and Joulin, Armand and Riedel, Sebastian and Grave, Edouard},
  journal={Journal of Machine Learning Research},
  volume={24},
  number={251},
  pages={1--43},
  year={2023}
}

@article{gao2023retrieval,
  title={Retrieval-augmented generation for large language models: A survey},
  author={Gao, Yunfan and Xiong, Yun and Gao, Xinyu and Jia, Kangxiang and Pan, Jinliu and Bi, Yuxi and Dai, Yixin and Sun, Jiawei and Wang, Haofen and Wang, Haofen},
  journal={arXiv preprint arXiv:2312.10997},
  volume={2},
  number={1},
  year={2023}
}

\appendix

\section{Data Filtering}
\label{appendix:data_filter}
In this section, we give an example of how questions from our benchmark are converted to text-only questions and how the Question Validity Filter and Answer Correctness Filter are performed in Table~\ref{tab:data_filtering}. This is an example of a prompt for questions generated from FMA. The prompt can be adapted to other questions generated from other datasets.

\begin{table}[h]
\centering
\caption{An example of the data filtering prompt.}
\begin{tabular}{p{0.95\linewidth}}
\hline
You are good at assessing the quality of a question. Please use the following criteria to determine the validity of the question:\\
a. If identifying the specific genre is essential to answer, output ``yes''. If the question can be answered directly from other information provided (without the genre), output ``no''.\\
b. The question is invalid if the question is vague or too general (e.g. refers only to ``an award-winning artist'')\\

After determining validity, check whether the provided answer correctly addresses the question. If correct, output ``no''; if not, output ``yes''.\\
\\

**Examples**\\
Example 1:\\
Genre: Hip-hop\\
Question: What award-winning artist, known for his lyrics and storytelling, released an album in 2022 that belongs to this genre?\\
Answer: Kendrick Lamar\\
Validity: no\\
Correctness: yes\\
$[$MORE EXAMPLES$]$\\
\\

**Your Turn**:\\
Question: {question}\\
Answer: {answer}\\
Ensure you strictly follow the above output format for ``Validity'' and ``Correctness'' (do not need to output ``Question'' and ``Answer'')\\
\hline
\end{tabular}
\label{tab:data_filtering}
\end{table}

\section{Evaluation Prompt}
\label{appendix:eval_prompt}

In this section, we present the evaluation prompt in Table~\ref{tab:evaluation_prompt}. It is used to assess the correctness of the models' answer and, at the same time, to categorize the incorrect responses into specific error types. 

\begin{table}[t]
\centering
\caption{Evaluation Prompt.}
\begin{tabular}{p{0.95\linewidth}}
\hline
You are given a question, an attribute of the audio, a ground truth answer, and a model answer. \\

Your task is to categorize the error in the model's answer into one of the following four types:\\
a) Reasoning Error:\\
- the model fails to understand the question or fails to apply correct reasoning to arrive at the answer;\\
- the error is not due to misrecognizing the audio or lack of external knowledge, but rather due to incorrect logic or interpretation;\\
b) Audio Processing Error:\\
- the question requires recognizing the attribute of the audio but the model recognizes it wrongly;\\
- if the model misinterprets the question, this is a **not** an Audio Processing Error, it **should be** Reasoning Error;\\
c) Knowledge Error: \\
- the question requires external knowledge beyond what is present in the audio, and the model gives the wrong answer due to lack of or incorrect knowledge;\\
d) Invalid Answer:\\
- the model gives no response or a nonsensical/incomplete output.\\
\\
Question:\\
\{question\}\\ 
\\
Audio Attribute:\\
\{audio\_attr\}\\
\\
Ground Truth:\\
\{gt\_answer\}\\
\\
Model Answer:\\
\{model\_answer\}\\
\\
Output only the error type (a,b,c or d). No additional explanation is needed. \\
\hline
\end{tabular}
\label{tab:evaluation_prompt}
\end{table}

\end{document}